\documentclass{article}
\usepackage{amsfonts}
\usepackage{amssymb}
\usepackage{amsmath}
\usepackage{amsthm}

\newcommand{\sh}{\text{sh}}
\newcommand{\ch}{\text{ch}}

\begin{document}

\title{\bf Non-trivial Minkowski backgrounds in f(T) gravity}

\author{Alexey Golovnev${}^{1}$,  Mar\'ia-Jos\'e Guzm\'an${}^{2,3}$\\
{\small ${}^{1}${\it Centre for Theoretical Physics, British University in Egypt,}}\\ 
{\small\it 11837 El Sherouk City, Cairo, Egypt,}\\
{\small agolovnev@yandex.ru}\\
{\small ${}^{2}${\it Departamento de F\'isica y Astronom\'ia, Facultad de Ciencias, Universidad de La Serena,}}\\
{\small\it Av. Juan Cisternas 1200, 1720236 La Serena, Chile,}\\
{\small ${}^{3}${\it Laboratory of Theoretical Physics, Institute of Physics, University of Tartu,}}\\ 
{\small {\it W. Ostwaldi 1, 50411 Tartu, Estonia}}\\
{\small maria.j.guzman.m@gmail.com}
}
\date{}

\maketitle

\begin{abstract}

Boosted and rotated tetrad backgrounds for the Minkowski space are studied in $f({\mathbb T})$ gravity. We perform Lorentzian perturbations at first order around non-trivial backgrounds and show that some Lorentz modes can exhibit non-trivial dynamics and can propagate in time. This remarkable feature gives evidence of additional mode(s) in the Lorentzian sector which have no precedent in Lorentz-violating modified gravities, however they can cast doubts onto even theoretical viability of these models.

\end{abstract}

\section{Introduction}

Modifications to Einstein's gravity are nowadays well motivated by the known problems in the $\Lambda$CDM standard model of cosmology, the need for an early accelerating inflationary phase in the early universe, and the quest for a quantum theory of gravity.
Amidst the many contenders intended to supersede general relativity, spacetimes based on a teleparallel framework appear as a conservative modification to the underlying geometry of space, that gives a genuine new modification of gravity. For a long time models of modified teleparallel gravity \cite{Ferraro:2006jd,Ferraro:2008ey,Bengochea:2008gz} were a very marginal direction in theoretical physics research. Nowadays the $f({\mathbb T})$ gravity is becoming widely used for cosmological model building and shows good performance at the level of linear cosmological perturbations in simplest models \cite{KoiGov, HHGZ}. It makes it very important to better understand the theoretical foundations of this modified gravity and to see whether it is viable, at least theoretically.

A natural path to test this class of theories is to study its perturbative behaviour. It is already known that at the level of linear perturbations no further gravitational wave modes compared to GR appear in $f({\mathbb T})$ equations of motion \cite{Bamba:2013ooa}, and those two which exist have the speed equal to the speed of light \cite{Cai:2018rzd}, therefore being the constraints of GW170817 and GRB170817A automatically satisfied. Note that in these works \cite{Bamba:2013ooa, Cai:2018rzd, Hohmann:2018jso} the Minkowski background is represented by the trivial tetrad given by a unit matrix in Cartesian coordinates. 
In Ref. \cite{Jimenez:2020ofm}, this trivial (diagonal) Minkowski background is perturbed arbitrarily (but only in the Lorentzian sector) and the action for $f({\mathbb T})$ gravity is computed up to fourth order, where the leading (4th) order term suggests 3 extra modes but only one of them dynamical.

It has been argued that the manifestation of at least one extra d.o.f. in $f({\mathbb T})$ gravity could appear under certain tetrad configurations satisfying $\mathbb T$ constant or time-dependent, and when certain Poisson brackets with  coefficients $F_{ab}$ in the Hamiltonian formalism are nonvanishing \cite{Ferraro:2018tpu}. The most general cases for the Hamiltonian formalism with $\mathbb T$ also depending on spatial coordinates are not covered by the Ref. \cite{Ferraro:2018tpu} and are treated in Refs. \cite{Li:2011rn,Blagojevic:2020dyq}. Non-perturbatively, the Hamiltonian analyses \cite{Ferraro:2018tpu,Li:2011rn,Blagojevic:2020dyq} contradict to each other but all show that at least one new dynamical mode exists.

But also in cosmological perturbations only the same number of dynamical modes as in GR shows up at the linear level \cite{KoiGov}. Given the evidence on the existence of extra modes, this could mean a possible strong coupling problem \cite{Jimenez:2020ofm}. Therefore it is very important to touch these modes explicitly and study their properties in order to make conclusions about viability of such models.

Since non-linear regimes in modified teleparallel gravities are quite complicated and cumbersome, it is worth looking at perturbations around other backgrounds. In particular one can study the Minkowski metric by using other corresponding tetrads that represent non-trivial backgrounds and its perturbative properties.

This manuscript is organized as follows. We introduce the reader to the theoretical framework of $f(\mathbb T)$ gravity and equations of motion in Sec. \ref{sec:gc}. We introduce the tetrad for a boosted Minkowski background in Sec. \ref{sec:bmb}, study Lorentzian perturbations around it in Sec. \ref{sec:lpbb}, and study the effect on perturbations by adding rotations in Sec. \ref{sec:rtb}. Our conclusions are in Sec. \ref{sec:concl}. 

\section{General considerations}
\label{sec:gc}

Let us consider a basis of tetrads $\mathbf{e}_a$ defined in the tangent space $T_p(M)$ of the 4-dimensional spacetime manifold $M$, and its associated co-tetrad basis $\mathbf{E}^{a}$ in the cotangent space $T^{\star}_p(M)$. Latin indices $a=0,\ldots,3$ will denote (co)tangent space indices. We can expand tetrad and cotetrad in a coordinate basis as $\mathbf{e}_a = e^{\mu}_a \partial_{\mu}$, $\mathbf{E}^{a}=E^{a}_{\mu}dx^{\mu}$, whose components satisfy completeness relations $E^{a}_{\mu}e_a^{\nu}=\delta^{\nu}_{\mu}$. The common metric from general relativity is retrieved by defining the norm of the the tetrads in terms of the Minkowski metric, that is
\begin{equation}
e^{\mu}_a e^{\nu}_b g_{\mu\nu} = \eta_{ab}, \qquad g_{\mu\nu} = E^{a}_{\mu} E^{b}_{\nu} \eta_{ab},
\end{equation}
where the metric is recovered in the second identity, and our signature convention is $\eta_{ab}=(1,-1,-1,-1)$.

We will work in pure tetrad formulation of $f({\mathbb T})$ gravity since it is equivalent to the covariant version with a flat spin connection \cite{GKS,Golovnev:2018red}, and is easier to deal with in concrete calculations because it has less variables. This assumption allows to write the connection for a flat spacetime as 
\begin{equation}
\label{conn}
\Gamma^{\alpha}_{\mu\nu}=e_a^{\alpha}\partial_{\mu}E^a_{\nu},  
\end{equation}
which is called Weitzenb\"{o}ck connection. This choice vanishes the Riemann tensor, but not the torsion tensor, which is
\begin{equation}
T^{\alpha}_{\hphantom{\alpha}\mu\nu}=\Gamma^{\alpha}_{\mu\nu}-\Gamma^{\alpha}_{\nu\mu}  = e^{\alpha}_a (\partial_{\mu} E^{a}_{\nu}-\partial_{\nu}E^{a}_{\mu}). 
\end{equation}
It will be useful to define the torsion vector $T_{\mu}= T^{\alpha}_{\hphantom{\alpha}\mu\alpha}$ and the torsion scalar ${\mathbb T}= T_{\alpha\mu\nu}S^{\alpha\mu\nu}$ where the so-called superpotential is given by
\begin{equation}
S_{\alpha\mu\nu}=\frac12\left(K_{\mu\alpha\nu}+g_{\alpha\mu}T_{\nu}-g_{\alpha\nu}T_{\mu} \right)
\end{equation}
with the contortion tensor being
\begin{equation*}
K_{\alpha\mu\nu}=g_{\alpha\beta}\left(\Gamma^{\beta}_{\mu\nu}-\mathop{\Gamma^{\beta}_{\mu\nu}}\limits^{(0)}\right)=\frac12 \left(T_{\alpha\mu\nu}+T_{\nu\alpha\mu}+T_{\mu\alpha\nu}\right).
\end{equation*}
Quantities written with $(0)$ on top are computed with the usual Levi Civita connection.

Finally we have all the tools that allow defining the $f({\mathbb T})$ gravity action
\begin{equation}
\label{eq:fT}
S = \dfrac{1}{2\kappa} \int d^{4}x E f({\mathbb T}),
\end{equation}
where $\kappa = 8 \pi G$ and $E=\text{det}(E^{a}_{\mu})$. In the limit $f({\mathbb T}) = {\mathbb T}$, the action \eqref{eq:fT} reduces to the teleparallel equivalent of general relativity (TEGR) action. TEGR is equal to GR due to the well-known relation between scalar curvatures of the two connections \cite{Golovnev:2018red}:
\begin{equation}
\label{telepid}
0=R(\Gamma)=R(\mathop\Gamma\limits^{(0)})+2 \mathop\bigtriangledown\limits^{(0)}{\vphantom{\bigtriangledown}}_{\mu}T^{\mu}+ \mathbb T.
\end{equation}
It shows that the TEGR action is equivalent to the GR one modulo a surface term. However, for a general  $f({\mathbb T})$ function, the resulting dynamics is drastically different from, for instance, $f(R)$ gravity. Our aim is to study linear perturbations to the Minkowski spacetime represented by non-trivial background tetrads.

\subsection{Minkowski space}

The study of perturbations to Minkowski spacetime is a basic consistency test for any gravitational theory. For $f({\mathbb T})$ gravity one can prove that the Minkowski solution is obtained with the trivial tetrad $E^a_{\mu}=\delta^a_{\mu} = \text{diag}(1,1,1,1)$. It is straightforward to show that such solution corresponds to a vanishing torsion scalar, that is ${\mathbb T}=0$ case. However, this tetrad is indeed a solution in vacuum if and only if $f(0)=0$, as it is shown later. In general, any solution with $\mathbb T = \text{const}$ is a solution of the equations of motion in $f(\mathbb T)$ gravity in vacuum (although a redefinition of gravitational and cosmological constants might be necessary) \cite{Ferraro:2011ks}.

Any Lorentz transformation of the tetrad preserves the metric, however a peculiarity of $f({\mathbb T})$ gravity is that if the Lorentz transformation depends on spacetime coordinates, a Lorentz transformed tetrad generically breaks the fulfillment of equations of motion, since the model is locally Lorentz breaking. However in some cases it might still be a solution (for instance, when one sticks to a member of the group of remnant symmetries \cite{Ferraro:2014owa} ). We will see that the same ${\mathbb T}=0$ condition is satisfied by the trivial tetrad boosted in one fixed direction (and possibly rotated in the orthogonal plane).

Since the equations in vacuum can be written \cite{KoiGov} as
\begin{equation}
\label{eom}
f_T({\mathbb T})\mathop{G_{\mu\nu}}\limits^{(0)} 
+ \frac12 \left
[\vphantom{f^A_B}f(\mathbb T)-f_T(\mathbb T){\mathbb T}\right]g_{\mu\nu}
+ 2f_{TT}(\mathbb T) S_{\mu\nu\alpha} \partial^{\alpha}{\mathbb T} = 0
\end{equation}
we see that only the third term with $f_{TT}$ term contributes to the antisymmetric part of the equations of motion. (Note that our superpotential $S$ is different by a $\frac12$ factor from many other works \cite{KoiGov,GKS,Golovnev:2018red}.) Also we see that any tetrad with Minkowski metric and ${\mathbb T}=0$ is a vacuum solution of every $f({\mathbb T})$ gravity with $f(0)=0$.

Moreover, in case of purely Lorentz perturbations, the only contribution to the perturbed equations is either from the
non-symmetric term or from the change of the argument in the scalar coefficients depending on $\mathbb T$ (i.e. $f_T(\mathbb T)$ in front of the Einstein tensor, and $f(\mathbb T)$ and $f_T(\mathbb T)$ in front of $g_{\mu\nu}$ ). The change of coefficients at linear order is only important in our case for the middle term (proportional to $g_{\mu\nu}$) in (\ref{eom}), which is the only one to exist at the zeroth order (and therefore $f(0)=0$ is required for existence of $\mathbb T=0$ Minkowski solutions).

Looking at the background superpotential (see \eqref{Scomp} in next section), we also see that the symmetric and antisymmetric parts of the $f_{TT}$ term coincide at linear order. It means that, once the antisymmetric equations are solved, the only terms of equation (\ref{eom}) which matter are the first two ones. 
Therefore, the linear metric perturbations behave the same way as in GR. The Lorentzian sector around the trivial tetrad fully disappears at linear level, therefore it is reasonable to study this sector around another tetrad representation of Minkowski space which is a Lorentz transformation of the trivial one, if one wants to find dynamics for Lorentzian modes.

Note that the antisymmetric part of this equation of motion can be considered as six equations for the four components of the gradient of the torsion scalar. If at least four of them are linearly independent then it would require that ${\mathbb T}=\text{const}$. It might mean that ${\mathbb T}\neq \text{const}$ solutions are fine-tuned in terms of superpotential properties. And we will also see that the cases with less independent equations have strange unusual properties of linear perturbations.

\section{Boosted Minkowski background}
\label{sec:bmb}

We consider the trivial tetrad, which corresponds to the unit matrix in Cartesian coordinates $E^b_{\mu}=\delta^{b}_{\mu}$, and we apply a boost in one direction with a Lorentz matrix $\Lambda^{a}_{b}$. Let us choose the direction of the boost in the $x$ coordinate, as our results are independent of the chosen direction. The boosted tetrad $E^{a}_{\mu}$ and inverse boosted tetrad $e^{\mu}_{a} $ are given by
\begin{equation}
\label{boosttetr}
E^{a}_{\mu} \equiv \Lambda^{a}_{b} \delta^{b}_{\mu} = \left( 
\begin{array}{cccc}
   \ch(\lambda)  & \sh(\lambda) & 0 & 0  \\
   \sh(\lambda)  & \ch(\lambda) & 0 & 0 \\
   0 & 0 & 1 & 0 \\
   0 & 0 & 0 & 1
\end{array}
\right), \qquad  e^{\mu}_{a} = (\Lambda^{-1})^{b}_{a} \delta^{\mu}_{b} = \left(
\begin{array}{cccc}
\ch(\lambda) & - \sh(\lambda) & 0 & 0 \\
-\sh(\lambda) & \ch(\lambda) & 0 & 0 \\
0 & 0 & 1 & 0 \\
0 & 0 & 0 & 1
\end{array} 
\right),
\end{equation}
where ch and sh are the hyperbolic trigonometric functions $\cosh$ and $\sinh$, and the boost parameter $\lambda$ is an arbitrary function of spacetime coordinates $\lambda=\lambda(t,x,y,z)$. Since this is a Lorentz rotation, the metric remains the Minkowski one. However, not every tetrad corresponding to Minkowski space would solve the equations of motion of $f({\mathbb T})$ (unless we go to the covariant formalism by adding a non-zero flat spin connection). To prove that equations (\ref{eom}) are satisfied for our boosted tetrad, it is enough to see that $\mathbb T=0$.

It is relatively easy to calculate the torsion scalar $\mathbb T$ for our background tetrad by hand. Indeed, since the Lorentz rotation does not change the curvatures, neither the Levi Civita one nor the teleparallel which is always zero, it is enough to prove that $\partial_{\mu} T^{\mu}=0$, recall the relation \eqref{telepid}. We see for the boosted background tetrad (\ref{boosttetr}) that
\begin{equation}
\partial_{\nu} E^{a}_{\mu} = \left( 
\begin{array}{cccc}
   \sh(\lambda)  & \ch(\lambda) & 0 & 0  \\
   \ch(\lambda)  & \sh(\lambda) & 0 & 0 \\
   0 & 0 & 0 & 0 \\
   0 & 0 & 0 & 0
\end{array}
\right)\partial_{\nu}\lambda,
\end{equation}
where the matrix describes the structure in indices $a$ and $\mu$. After that we calculate $T_{\mu}=-e^{\nu}_a \partial_{\nu} E^{a}_{\mu}=(\lambda^{\prime}, \dot\lambda, 0, 0)$ where $\dot{()}=\frac{d}{dt}$ and $()^{\prime}=\frac{d}{dx}$, so that the vanishing of the  divergence $\partial_{\mu}T^{\mu}$ becomes obvious, which consequently implies $\mathbb T = 0$, by virtue of \eqref{telepid}.

The only non-vanishing connection coefficients are $\Gamma^0{}_{\nu 1}=\Gamma^1{}_{\nu 0}=\partial_{\nu}\lambda$, and the background superpotential reads
\begin{equation}\label{Scomp}
\begin{split}
S_{ytx} & = -S_{yxt} = \lambda_y/2, \\
S_{yty} & = -S_{yyt} = -\lambda_x/2, \\
S_{yxy} & = -S_{yyx} = -\lambda_t/2, \\
S_{ztx} & = -S_{zxt} = \lambda_z/2, \\
S_{ztz} & = -S_{zzt} = -\lambda_x/2, \\
S_{zxz} & = -S_{zzx} = -\lambda_t/2,
\end{split}
\end{equation}
or without the factor $\frac12$ using the other definition of $S$. Note that for any two values of the first two indices only one position of them produces a non-zero component of $S$, for example $S_{ytx}\neq 0$ but then $S_{tyx}=0$, therefore its symmetric and antisymmetric parts are equal to each other.

This property of the superpotential means that even for linear perturbations in our case once we have solved the antisymmetric part of equations (\ref{eom}) the $f_{TT}$ term stops influencing also the symmetric part. Then the first term does not depend on perturbations of $\mathbb T$ at the linear order since the Einstein tensor is zero for Minkowski space. Also we see that $\delta f=f_T(0) \mathbb T$, and since $f(0)=0$ the second term of (\ref{eom}) does not appear at all before the quadratic order. To summarise, the symmetric part of equations takes the GR form $f_T(0)\mathop{G_{\mu\nu}}\limits^{(0)}=0$. Therefore metric and Lorentz perturbations completely decouple in the linear approximation. This means that metric perturbations are then equivalent to the case of GR, and purely Lorentz perturbations can be studied separately, which is what we do in the next Section. (If this property of $S$ was not the case, then the Lorentz perturbations would make a source term for the usual gravitons.)

If one is only interested in antisymmetric equations, the  coefficients in front of the derivatives of $\mathbb T$ can be easily found without calculating the superpotential itself, since
\begin{equation}
A_{\mu\nu\alpha}\equiv 2(S_{\mu\nu\alpha}-S_{\nu\mu\alpha})=T_{\alpha\mu\nu}+g_{\alpha\mu}T_{\nu}-g_{\alpha\nu}T_{\mu}.
\end{equation}
This gives
\begin{equation}
\label{Aback}
\begin{split}
A_{tyx} =  -A_{ytx} & = \lambda_y, \\
A_{tzx} =  - A_{ztx} & = \lambda_z, \\
A_{xyt} =  -A_{yxt} & = \lambda_y, \\
A_{xzt} =  -A_{zxt} & = \lambda_z, \\
A_{tba} =  - A_{bta} & = \delta_{ba} \lambda_x, \\
A_{xba} =  -A_{bxa} & = \delta_{ab} \lambda_t
\end{split}
\end{equation}
where $a$ and $b$ can be equal to $y$ or $z$, but not $x$.

Before considering perturbations, let us mention questions of covariant formulation. Some person would say that the boost of the tetrad must be accompanied by the appropriate transformation of the spin connection. This is possible to do, however not only this boost (which is a remnant symmetry), but absolutely any arbitrary Lorentz transformation would then leave it a solution. And moreover, it would be totally the same solution, just written in another reference frame, while what we really want is to have another solution for changing the properties of perturbations.

Let us illustrate it. In the covariant approach we would obtain the following spin connection for the boosted tetrad:
\begin{equation}
\omega^a_{\hphantom{a}\mu b}=-(\partial_{\mu}\Lambda^a_c)({\Lambda^{-1}})^c_b=-\left( 
\begin{array}{cccc}
   0  & 1 & 0 & 0  \\
   1  & 0 & 0 & 0 \\
   0 & 0 & 0 & 0 \\
   0 & 0 & 0 & 0
\end{array}
\right)\partial_{\mu}\lambda,
\end{equation}
which in the covariant definition of the connection $\Gamma^{\alpha}_{\mu\nu}=e_a^{\alpha}\left(\partial_{\mu}E^a_{\nu}+\omega^a_{\hphantom{a}\mu b}E^b_{\nu}\right)$ simply cancels the only non-vanishing connection components we had (since $e\omega E=\Lambda^{-1}\omega\Lambda$ is equal to $\omega$ itself) making $\Gamma^{\alpha}{}_{\mu\nu}=0$. Therefore the boosted tetrad with this spin connection has the torsion tensor equal to zero so that, the same as around the trivial pure tetrad Minkowski solution, the perturbations come only in the second order of $\delta \mathbb T$ and the linearised theory again behaves like TEGR=GR. 

This is of course true for any Lorentz transformation if it is done on both tetrad and spin connection. Indeed, the non-invariance of the pure tetrad formulation is due to the new term of the form $e\Lambda^{-1}(\partial \Lambda) E$ in the connection coefficient. However, in the covariant definition of the spacetime connection the second term then changes from zero to $e\Lambda^{-1}\omega \Lambda E=-e\Lambda^{-1}(\partial\Lambda)\Lambda^{-1} \Lambda E$ which cancels the new contribution from the pure tetrad term. Therefore it stays the same solution, with both metric and torsion unchanged.

On the other hand, our boosted tetrad with zero spin connection is genuinely a new solution made possible by the remnant symmetry, and we will see new linear perturbation properties for it. And this also proves that the trivial tetrad allows not only for the zero spin connection. Indeed, starting from the boosted tetrad and zero spin connection we can transform back to the diagonal tetrad and
\begin{equation}
\omega^a_{\hphantom{a}\mu b}=-\partial_{\mu}({\Lambda^{-1}})^a_c \Lambda^c_b=\left( 
\begin{array}{cccc}
   0  & 1 & 0 & 0  \\
   1  & 0 & 0 & 0 \\
   0 & 0 & 0 & 0 \\
   0 & 0 & 0 & 0
\end{array}
\right)\partial_{\mu}\lambda.
\end{equation}
It gives to the diagonal tetrad the same connection coefficients as we had above instead of zero ones. Then we have again $\mathbb T=0$ but the perturbations as in the next Section.

\section {Lorentzian perturbations around the boosted background}
\label{sec:lpbb}

If we are truly devoted to finding something new at the linear level in Minkowski background, we should think about linear perturbations of the tetrad in purely Lorentzian sector, without changing the metric. At the linearised level, we take
\begin{equation}\label{linl}
\delta E^{a}_{\mu} = s_{ab} E^{b}_{\mu}
\end{equation}
with an arbitrary Lorentz generator 
\begin{equation}\label{Lgen}
s_{ab} = \left( 
\begin{array}{cccc}
0 & s_{01} & s_{02} & s_{03} \\
s_{01} & 0 & s_{12} & s_{13} \\
s_{02} & -s_{12} & 0 & s_{23} \\
s_{03} & -s_{13} & -s_{23} & 0
\end{array}
\right)
\end{equation}
which at the linear level leaves the metric intact. In Eq. \eqref{linl}, and from the Lorentz group viewpoint, one of the indices of this matrix has to be thought of as an upper one, that is $\delta E^{a}_{\mu} = s^{a}{}_{b} E^{b}_{\mu}$, however handling the specific components is easier when both indices are down, as defined in \eqref{Lgen}. 

Explicitly, the perturbed tetrad to first order can be written as 
\begin{equation}
\label{1stotetr}
\delta E^{a}_{\mu} = \left( 
\begin{array}{cccc}
   \sh(\lambda)s_{01}  & \ch(\lambda)s_{01} & s_{02} & s_{03}  \\
   \ch(\lambda)s_{01}  & \sh(\lambda)s_{01} & s_{12} & s_{13} \\
   \ch(\lambda)s_{02}-\sh(\lambda)s_{12} & \sh(\lambda)s_{02} - \ch(\lambda)s_{12} & 0 & s_{23} \\
   \ch(\lambda)s_{03} - \sh(\lambda)s_{13} & \sh(\lambda)s_{03} - \ch(\lambda)s_{13} & -s_{23} & 0
\end{array}
\right).
\end{equation}

One can calculate the perturbation of the  torsion scalar by hand using 
\begin{equation}
{\mathbb T}=-2\partial_{\mu}T^{\mu}=2\partial_{\mu}(\eta^{\mu\nu}e^{\alpha}_a\partial_{\nu}E^a_{\mu}),
\end{equation}
from which we will compute the first order approximation. The non-zero contribution comes only from the internal derivative acting on the Lorentz generator $s$, not on the background $E^{a}_{\mu}$, and it is enough then to take the components of $e^{\mu}_{a}$ at the background level. It means that the only relevant part of $e^{\alpha}_a\partial_{\nu}E^a_{\mu}$ is given by $\delta^{\alpha}_a ({\Lambda^{-1}})^a_b (\partial_{\nu}s^b_c) \Lambda^c_d \delta^d_{\mu}$ or $\Lambda^{-1}(\partial_{\nu}s)\Lambda$ in a beautiful index-free notation.  
The resulting expression for the linearised torsion scalar is
\begin{equation}
\label{Tpert}
\begin{split}
\dfrac{1}{2}\mathbb T & =  \lambda_{t}[\sh(\lambda)s_{03,z} - \ch(\lambda)s_{13,z} + \sh(\lambda)s_{02,y} - \ch(\lambda)s_{12,y} ]  \\
& + \lambda_{x}[\sh(\lambda)s_{13,z} -\ch(\lambda) s_{03,z} - \ch(\lambda)s_{02,y} + \sh(\lambda) s_{12,y} ]\\
& + \lambda_{y}[\ch(\lambda)s_{02,x} - \sh(\lambda)s_{12,x} - \sh(\lambda)s_{02,t} + \ch(\lambda)s_{12,t} ] \\
& + \lambda_{z}[\ch(\lambda)s_{03,x} - \sh(\lambda)s_{13,x} - \sh(\lambda)s_{03,t} + \ch(\lambda)s_{13,t} ].
\end{split}
\end{equation}
In this expression the overall sign depends on whether mostly plus or mostly minus convention for the metric signature is used.

The antisymmetric part of the equations of motion (\ref{eom}) has a very simple form
\begin{equation}
f_{TT} A_{\mu\nu\alpha}\partial^{\alpha} \mathbb T =0,
\end{equation}
and we assume that $f_{TT}\neq 0$.
At the linear level we have to take $A_{\mu\nu\alpha}$ at the background value (\ref{Aback}). It gives four equations in our case for our particular boost:
\begin{equation} \label{linearEqs}
\begin{split}
- \lambda_y \partial_x \mathbb T + \lambda_x \partial_y \mathbb T & = 0,\\
- \lambda_z \partial_x \mathbb T + \lambda_x \partial_z \mathbb T & = 0,\\
- \lambda_y \partial_t \mathbb T + \lambda_t \partial_y \mathbb T & = 0, \\
- \lambda_z \partial_t \mathbb T + \lambda_t \partial_z \mathbb T & = 0,
\end{split}
\end{equation}
considering the linear perturbation (\ref{Tpert}) on $\mathbb T$.
Only the last two equations of \eqref{linearEqs} will include 2nd order time derivatives of the modes $s_{ab}$. We also notice that these time derivatives are always accompanied by factors where derivatives of $\lambda$ are only on $y$ or $z$ coordinates, so in order to have dynamics for the Lorentz modes it is required that $\lambda_y$ or $\lambda_z$ are non-zero. Note also that only three equations are independent since the first two and the last two equations imply the same relation
\begin{equation}
\dfrac{\partial_y \mathbb T}{\lambda_y} = \dfrac{\partial_z \mathbb T}{\lambda_z}
\end{equation}
between $\partial_y \mathbb T$ and $\partial_z \mathbb T$.

If all four equations for the gradient of $\mathbb T$ in \eqref{linearEqs} were linearly independent, it would require that ${\mathbb T}=\text{const}$, even at the level of perturbations. Since $\mathbb T$ depends only on first derivatives of $s_{ab}$, it would amount to four second order differential equations ($\partial_{\mu}{\mathbb T}=0$), only one with second time derivative ($\partial_t {\mathbb T}=0$) which possibly gives hints to one new dynamical degree of freedom with some of the six $s_{ab}$ variables undetermined (remnants of gauge freedom). However when the equations are less restrictive, and leave the partial derivatives of $\mathbb T$ free in one or more directions, then the resulting model can present a strange behaviour as we will see below. 

\subsection{A special case}
\label{specialcase}

Let us consider a simple example with $$\lambda=\lambda(z)$$ assuming that $\lambda_z \neq 0$. We can easily find the antisymmetric part of the equations of motion at the linear level since 
\begin{equation}
\dfrac{1}{2}\mathbb T =   \lambda_{z}[\ch(\lambda)s_{03,x} - \sh(\lambda)s_{13,x} - \sh(\lambda)s_{03,t} + \ch(\lambda)s_{13,t} ] 
\end{equation}
and the only two non-trivial equations for the antisymmetric part
\begin{equation}
-\lambda_z \partial_x \mathbb T = 0, \ \ \ \ \ -\lambda_z \partial_t \mathbb T = 0
\end{equation}
become
\begin{equation}\label{lineqsmodes}
\begin{split}
 \ch(\lambda) \ddot{s}_{13} - \sh(\lambda) \ddot{s}_{03} -\sh(\lambda) \dot{s}^{\prime}_{13} + \ch(\lambda) \dot{s}^{\prime}_{03} & =  0, \\
 \ch(\lambda) s^{\prime\prime}_{03} - \sh(\lambda) s^{\prime\prime}_{13} - \sh(\lambda) \dot{s}^{\prime}_{03} + \ch(\lambda)\dot{s}^{\prime}_{13} & =  0.
 \end{split}
\end{equation}
Only Lorentz transformations in $tz$ and $xz$ planes are the infinitesimal perturbations that enter the linear equations \eqref{lineqsmodes} ($s_{03}$ and $s_{13}$, respectively), others being totally free.

This system of equations can be greatly simplified defining the new variables 
\begin{equation}\label{eqmodes}
\begin{split}
\xi & =  \sh(\lambda)s_{03}-\ch(\lambda)s_{13}, \\
\eta & = -\ch(\lambda)s_{03}+\sh(\lambda)s_{13}.
\end{split}
\end{equation}
Since $\sh(\lambda)=\ch(\lambda)$ is satisfied only when $\lambda \rightarrow \infty$, then this redefinition is always well-defined. The system of equations becomes
\begin{equation}
\begin{split}
\ddot{\xi} + \dot{\eta}^{\prime} & =  0, \\
\eta^{\prime\prime}+\dot{\xi}^{\prime} & =  0.
\end{split}
\end{equation}

Both equations simultaneously imply that the combination $\dot{\xi}+\eta^{\prime}$ cannot depend on $t$ and $x$. This means that
\begin{equation}
\dot{\xi}+\eta^{\prime} = c_1(y,z).
\end{equation}
On the other hand, $\eta$ can be chosen arbitrarily: $\eta=\eta(t,x,y,z)$, such that the previous equation implies
\begin{equation}
\dot{\xi} = c_1(y,z) - \eta^{\prime}(t,x,y,z),
\end{equation}
and when integrating with respect to time, we get the general solution
\begin{equation}
\xi = c_1(y,z) t - \int_0^t d\tau\cdot \eta^{\prime}(\tau,x,y,z) - c_2(x,y,z)
\end{equation}
which is a mode that has dependence also on $t$ and $x$, except for the first term that does not depend on $x$. 

We can informally say that this is ``a bit less than one degree of freedom''. Out of the two Cauchy data needed for $c_1$ and $c_2$, one of the initial conditions depends only on two coordinates, not on all three. It comes from the fact that in this case the torsion scalar cannot change along the direction of the background boost. If $\dot\xi +\eta^{\prime}$ had to be constant in all directions, we would have $c_1$ constant and precisely one half degree of freedom (one Cauchy datum $c_2$) plus one half global ($c_1$) freedom. 

The reported behaviour seems problematic and puts physical viability of the theory at question. One half degree of freedom is not an impossible physical scenario though, even in a well-behaved Hamiltonian formalism, a constraint might get a non-zero Poisson bracket with itself if proportional to the antisymmetric function $\delta^{\prime}$ and thus giving an odd number of second class constraints. In particular, in Horava gravity models with half-integer number of degrees of freedom were found, see \cite{Henneaux:2009zb,Sotiriou:2010wn}. A simpler example that also exhibits this behavior can be found for chiral bosons \cite{Henneaux:2009zb}. 
  
Our finding that the Cauchy data for the function $c_1$ depend on a less number of spacetime coordinates (neither all of them, nor none of them) is probably a worse behaviour. We are not aware of similar cases in other gravity or physical models. At the level of partial differential equations, it represents systems of the  $V^{\mu}\partial_{\mu}\phi=0$ kind with a background vector $V^{\mu}$, which restricts  perturbations of $\phi$ only along some directions.

\subsection{Another parametrisation of perturbations}

Another way to parametrise perturbations is to first perturb the trivial tetrad and then apply the matrix $\Lambda^{a}_{
b}$. It means that at the linear level $\delta E=\Lambda\cdot s$, or $\delta E^a_{\mu}=\Lambda^a_b s^b_c \delta^c_{\mu}$ in components. The calculations become a bit more cumbersome, and are better done on a computer, but yield a simpler answer for the torsion scalar:
\begin{equation}\label{simpleT}
\begin{split}
\frac12 {\mathbb T} & =  \dfrac{\partial \lambda}{\partial z}\left( \dfrac{\partial s_{03} }{\partial x} + \dfrac{\partial s_{13} }{\partial t} \right) + \dfrac{\partial \lambda}{\partial y} \left(\dfrac{\partial s_{12} }{\partial t} + \dfrac{\partial s_{02} }{\partial x} \right) \\
& - \dfrac{\partial \lambda}{\partial t}\left( \dfrac{\partial s_{13} }{\partial z} + \dfrac{\partial s_{12} }{\partial y} \right) - \dfrac{\partial \lambda}{\partial x} \left( \dfrac{\partial s_{03}}{\partial z} + \dfrac{\partial s_{02}}{\partial y} \right).
\end{split}
\end{equation}

This is actually related to our change of variables \eqref{eqmodes} for the simple example with $\lambda(z)$. Our initial presentation for the tetrad was $s \cdot  \Lambda$ (in terms of matrix multiplication). Now we want $\Lambda \cdot  \tilde s$ for the same perturbation instead. Therefore $\tilde s=\Lambda^{-1}\cdot  s \cdot \Lambda$. This is precisely the change of variables which makes the torsion scalar in \eqref{simpleT} and therefore the equations simpler than the more traditional approach firstly adopted.

In our special case we also see the geometrical meaning of the variable ($\xi$ in  previous variables) which has the second time derivative in the equations. For $\lambda=\lambda(z)$, it is just $s_{13}$ in the new variables, the $xz$ rotation, in terms of perturbation to the trivial tetrad before transforming everything by $\Lambda$.

\section{Adding rotation to boost}
\label{sec:rtb}

Actually one can easily see that a more general tetrad, namely
\begin{equation}
\Lambda^{a}_{b} = \left( 
\begin{array}{cccc}
   \ch(\lambda)  & \sh(\lambda) & 0 & 0  \\
   \sh(\lambda)  & \ch(\lambda) & 0 & 0 \\
   0 & 0 & \cos(\psi) & -\sin(\psi) \\
   0 & 0 & \sin(\psi) & \cos(\psi)
\end{array}
\right)
\end{equation}
also gives $\mathbb T=0$ and therefore solves the equations of vacuum $f({\mathbb T})$ if $f(0)= 0$. As before, it can be seen by finding that $T_{\mu}=-e^{\nu}_a \partial_{\nu} E^{a}_{\mu}=(\lambda_x, \lambda_t, \psi_z, -\psi_y)$ obviously has zero divergence.

For perturbations (in the alternative parametrisation) it gives
\begin{small}
\begin{equation}
\begin{split}
\frac12 {\mathbb T} & = \dfrac{\partial \lambda}{\partial z}\left( \dfrac{\partial s_{03} }{\partial x} + \dfrac{\partial s_{13} }{\partial t} \right) + \dfrac{\partial \lambda}{\partial y} \left(\dfrac{\partial s_{12} }{\partial t} + \dfrac{\partial s_{02} }{\partial x} \right)  - \dfrac{\partial \lambda}{\partial t}\left( \dfrac{\partial s_{13} }{\partial z} + \dfrac{\partial s_{12} }{\partial y} \right) - \dfrac{\partial \lambda}{\partial x} \left( \dfrac{\partial s_{03}}{\partial z} + \dfrac{\partial s_{02}}{\partial y} \right) \\
&  +\dfrac{\partial \psi}{\partial z} \left( \dfrac{\partial s_{12}}{\partial x} + \dfrac{\partial s_{02}}{\partial t} \right) - \dfrac{\partial \psi}{\partial y}\left( \dfrac{\partial s_{13}}{\partial x}+ \dfrac{\partial s_{03}}{\partial t} \right) - \dfrac{\partial \psi}{\partial x}\left( \dfrac{\partial s_{12}}{\partial z} - \dfrac{\partial s_{13}}{\partial y} \right) - \dfrac{\partial \psi}{\partial t}\left( \dfrac{\partial s_{02}}{\partial z} - \dfrac{\partial s_{03}}{\partial y} \right) 
\end{split}
\end{equation}
\end{small}
for the linearised torsion scalar (again no dependence on $s_{01}$ or $s_{23}$ since those components of $s$ would only change the arbitrary $\lambda$ and $\psi$ respectively) and the coefficients of the antisymmetric part of equations as
\begin{equation}
\begin{split}
A_{tyt} & = \psi_z, \qquad \ A_{tyx}  =  -\lambda_y, \qquad A_{tyy}  = \lambda_x, \qquad \ \ A_{tyz}  =  \psi_t, \\
A_{tzt} & =  -\psi_y,\ \quad A_{tzx}  =  -\lambda_z, \qquad A_{tzy}  =  -\psi_t, \qquad A_{tzz}  =  \lambda_x, \\
A_{xyt} & =  \lambda_y, \qquad A_{xyx}  =  -\psi_z, \qquad A_{xyy}  =  \lambda_t, \qquad \ \ A_{xyz}  =  \psi_x, \\
A_{xzt} & =  \lambda_z, \qquad A_{xzx}  =  \psi_y, \qquad \ \ A_{xzy}  =  -\psi_x, \qquad A_{xzz}  =  \lambda_t.
\end{split}
\end{equation}
It again amounts to four equations (no $xt$ and $yz$ components). They take the following form:
\begin{equation}
\begin{split}
-\psi_z {\mathbb T}_t-\lambda_y {\mathbb T}_x+\lambda_x {\mathbb T}_y+\psi_t {\mathbb T}_z & = 0,\\
\psi_y {\mathbb T}_t-\lambda_z {\mathbb T}_x-\psi_t  {\mathbb T}_y+\lambda_x {\mathbb T}_z & = 0,\\
-\lambda_y {\mathbb T}_t-\psi_z {\mathbb T}_x+\lambda_t {\mathbb T}_y+\psi_x {\mathbb T}_z & = 0,\\
-\lambda_z {\mathbb T}_t+\psi_y {\mathbb T}_x-\psi_x {\mathbb T}_y+\lambda_t {\mathbb T}_z & = 0.
\end{split}
\end{equation}
Now this system is generically non-degenerate. For example, in the case of $\psi(t,z)$ and $\lambda(z)$ we get ${\mathbb T}=\text{const}$ as the only solution.  

If to take $\psi = \text{const}$, the system of equations reduces to the previous case of the pure boost. In case of arbitrary $\psi$ and $\lambda=\text{const}$ we get the background with purely rotated tetrad. One can assume then that $\psi=\psi(z)$ which would result in ${\mathbb T}={\mathbb T}(y,z)$ and the situation is then very similar to the special case we considered in Section \ref{specialcase} for the boosted background tetrad.

\section{Discussion and conclusions}
\label{sec:concl}

The properties of Minkowski space in $f(\mathbb T)$ gravity have been studied in the past  \cite{Bamba:2013ooa, Cai:2018rzd, Hohmann:2018jso} but the Minkowski background is represented by a trivial diagonal tetrad, and no new propagating modes have been found. In Ref. \cite{Jimenez:2020ofm}, this diagonal Minkowski background has been perturbed arbitrarily only in the Lorentzian sector and replaced back in the action for $f({\mathbb T})$ gravity. Up to fourth order, the leading 4th order term suggests 3 extra modes but only one dynamical. One natural way of extending these works is considering that modified Minkowski backgrounds could exhibit additional d.o.f. at lower orders in perturbation theory. 

Non-perturbative approaches on the d.o.f., for instance the Hamiltonian formalism performed in \cite{Ferraro:2018tpu}, predict one extra degree of freedom for $f(\mathbb T)$ gravity in a Minkowski or FLRW  background. Although it has been pointed out that the computation of the constraint algebra is substantially wrong \cite{Blagojevic:2020dyq}, and indeed the analysis can not be considered reliable for the most general case $\partial_i \mathbb T \neq 0$, the analysis in  \cite{Ferraro:2018tpu} seems to describe the important cases for which $\partial_i \mathbb T = 0$ (spatially constant torsion scalar) and still predict an additional d.o.f.. These claims are also contradicted by \cite{Blagojevic:2020dyq}, however as the present work and previous one suggest \cite{Jimenez:2020ofm}, the claims on the non-appearance of extra modes in Minkowski spacetime in \cite{Blagojevic:2020dyq} are wrong. A possible explanation on the discrepancy is that in \cite{Blagojevic:2020dyq} the authors consider that cases with $\partial_i \mathbb T = 0$ are equivalent to GR with shifted gravitational and cosmological constants, therefore they conclude that such cases should have the same d.o.f. as GR. This is true at the background level but, however, can not be possibly true at a perturbative level.

In \cite{Ferraro:2018tpu}, the appearance of one extra d.o.f. is attributed to the non-vanishing of the coefficients $F_{ab} = 4 E \partial_i E^c{}_j \left( e_{[b}{}^0 e_{a]}{}^i e_{c}{}^j + e_{[b}{}^i e_{a]}{}^j e_{c}{}^0 + e_{[b}{}^j e_{a]}{}^0 e_{c}{}^i \right)$, and their vanishing or non-vanishing  changes the rank of the matrix of PB brackets among constraints, therefore the counting of d.o.f.. It is clear that for the diagonal trivial Minkowski tetrad such coefficients are zero, but that is not the case for the non-trivial Minkowski backgrounds shown in this article. If we were to believe the results in \cite{Ferraro:2018tpu}, then the reason we observe additional d.o.f. at lower orders in perturbation theory (at 1st order) is because background tetrads with non-vanishing $F_{ab}$ are more likely to exhibit them, in this sense, such Hamiltonian analysis can help to find non-trivial backgrounds on which the new modes are likely to appear, and could also have some saying on cosmological perturbations.

In this work we have derived equations of linear perturbations in a Lorentz rotated Minkowski background in $f({\mathbb T})$ theories. The perturbations in the Lorentzian sector do not trivialise and give new Cauchy data to be used. This is a novel result and it does not appear in other Lorentz violating-like modified gravities, to the best of our knowledge. This new ``Lorentzon'' mode, which in the example given in our special case in Sec.\ref{specialcase} is a  combination of an infinitesimal boost in $tz$ and rotation in $xz$ plane, gets dynamical equations of motion in the antisymmetric part and solves them for the particular choices exhibited there. One may wonder if such new modes could have some physical realisation, since they only cause oscillations in the tetrad and not in the metric. The consideration of coupling of matter with spin could give some answer to such concerns. 

Also an important issue is that any new modes in $f({\mathbb T})$ gravity refer to the strong coupling problem \cite{Jimenez:2020ofm} in cosmology \cite{KoiGov} and weak gravity. If new modes appear in a slightly varied background it means that they exist in non-linear corrections for the initial background. 

For the general case we also see that in cases when linearised equations for the components of the gradient of $\mathbb T$ are a degenerate system, the initial value problem acquires unusual features in what concerns the nature of free initial functions. This effect of a function depending on only some non-empty subset of spatial coordinates might be disappearing though (turning into a global constant) at higher order corrections because we saw that around more generic backgrounds the equations become non-degenerate and do require that ${\mathbb T}= \text{const}$. 

On top of the strong coupling, it is yet another indication of unreliability of linearised considerations in these models. If, on the contrary, the Cauchy datum with dependence on a non-full set of spatial coordinates is a real property surviving beyond just the lower approximations, then it is also very bad news for using this model as a physically viable one, for it is different from how we can usually describe nature. And the boosted backgrounds have these perturbation properties even being arbitrarily close to simple symmetric solutions which are supposed to be used for weak gravity and similarly for cosmological model building.

The local Lorentz invariance gets broken in $f({\mathbb T})$ gravity in a very specific way related to the surface term \cite{Ferraro:2018axk,Ferraro:2020tqk}, and it seems to make a very strange behaviour of variables when one departs from the simplest and most trivial solutions. Given the naive success of modified teleparallel gravities in cosmology \cite{HHGZ}, it calls for better study of many other types of models of this sort.

\section*{Acknowledgments}
The authors would like to thank R. Ferraro for helpful discussions in an early stage of this work, and to S. Bahamonde for constructive criticism in a previous version of this manuscript. M.J.G. was funded by FONDECYT-ANID postdoctoral grant 3190531.

\end{document}